\documentclass[a4paper,fleqn,usenatbib]{mnras}

\usepackage{mathptmx}

\usepackage[T1]{fontenc}
\usepackage{ae,aecompl}

\usepackage{graphicx}	
\usepackage{amsmath}	
\usepackage{amssymb}	

\usepackage{times}

\usepackage{url}

\newcommand{\teff}{$T_{\mathrm{eff}}$}
\newcommand{\muhz}{$\mu$Hz}
\newcommand{\numax}{$\nu_{\mathrm{max}}$}
\newcommand{\dnu}{$\Delta\nu$}
\newcommand{\dP}{$\Delta P$}
\newcommand{\msol}{M$_\odot$}
\newcommand{\rsol}{R$_\odot$}
\newcommand{\lsol}{L$_\odot$}
\newcommand{\kepler}{\textit{Kepler}}

\title[Seismic masses of evolved planet hosts]{Asteroseismic masses of retired planet-hosting A-stars using SONG}

\author[D. Stello et al.]{
Dennis~Stello$^{1,2,3}$
Daniel~Huber$^{4,2,5,3}$   
Frank~Grundahl$^{3}$       
James~Lloyd$^{6}$          
Mike~Ireland$^{7}$\and     
Luca~Casagrande$^{7}$      
Mads~Fredslund$^{3}$       
Timothy~R.~Bedding$^{2,3}$     
Pere~L.~Palle$^{8}$        
Victoria~Antoci$^{3}$\and  
Hans~Kjeldsen$^{3}$        
J\o rgen~Christensen-Dalsgaard$^{3}$ 
\\
$^{1}$School of Physics, University of New South Wales, NSW 2052, Australia\\
$^{2}$Sydney Institute for Astronomy (SIfA), School of Physics, University of Sydney, NSW 2006, Australia\\
$^{3}$Stellar Astrophysics Centre, Department of Physics and Astronomy, Aarhus University, DK-8000 Aarhus C, Denmark\\
$^{4}$Institute for Astronomy, University of Hawai i, 2680 Woodlawn Drive, Honolulu, HI 96822, USA\\
$^{5}$SETI Institute, 189 Bernardo Avenue, Mountain View, CA 94043, USA\\
$^{6}$Department of Astronomy, Cornell University, Ithaca, NY 14850, USA\\
$^{7}$Research School of Astronomy \& Astrophysics, Mount Stromlo Observatory, The Australian National University, ACT 2611, Australia\\
$^{8}$Instituto de Astrof sica de Canarias, E-38200 La Laguna, Tenerife, Spain
}

\date{Accepted XXX. Received YYY; in original form ZZZ}

\pubyear{2017}

\begin{document}
\label{firstpage}
\pagerange{\pageref{firstpage}--\pageref{lastpage}}
\maketitle

\begin{abstract}
To better understand how planets form, it is important to study planet
occurrence rates as a function of stellar mass. 
However, estimating masses of field stars is often difficult.  
Over the past decade, a controversy has arisen about the inferred 
occurrence rate of gas-giant planets around evolved intermediate-mass stars -- the so-called `retired A-stars'. 
The high masses of these red-giant planet hosts, derived using spectroscopic information and
stellar evolution models, have been called into question.
Here we address the controversy by determining the masses of eight
evolved planet-hosting stars using asteroseismology.  We compare the
masses with spectroscopic-based masses from the Exoplanet Orbit Database
that were previously adopted to infer properties of the exoplanets and
their hosts.  
We find a significant one-sided offset between the two sets of masses for
stars with spectroscopic masses above roughly 1.6\msol, suggestive of an 
average 15--20\% overestimate of the adopted spectroscopic-based masses.
The only star in our sample well below this mass limit is also the only
one not showing this offset.   
Finally, we note that the scatter across literature values of
spectroscopic-based masses often exceed their formal uncertainties, making
it comparable to the offset we report here.    
\end{abstract}




\section{Introduction}
One way to understand how planets form is to find relationships between
the occurrence rates of exoplanets and the fundamental properties of their host
stars, such as mass.   
The search for exoplanets has mainly been focused on cool main-sequence
stars below $\sim 1.4\,$\msol\ because their planets are easier to detect \citep{Johnson06}. 
To 
extend the mass range of exoplanet host targets, \citet{Johnson06} 
searched for planets around 
cool evolved intermediate-mass stars, dubbed retired A-stars\footnote{While
not all the retired A-stars are strictly speaking old A-stars (some are
less massive), we adopt this previously-dubbed `group-name' for
simplicity.}, which were once hotter main sequence stars, from which planet
occurrence rates could be inferred.
To estimate stellar mass, they used an isochrone grid-modelling
approach based on spectroscopic input observables ($\log\,g$, \teff, and
[Fe/H]).  From this, \citet{Johnson07a,Johnson07b} reported a general
increased planet occurrence rates but a paucity of planets in short-period
orbits.    

However, the mass estimates of the retired A-stars were subsequently
called into question by \citet{Lloyd11}, who argued it was statistically
unlikely that the sample, which he extracted from the Exoplanet Orbit
Database (EOD) \citep{Wright11}, would include so many relatively massive
stars, given their location in the HR-diagram.  This led to further
investigations by \citet{Johnson13}, \citet{Lloyd13}, and \citet{SchlaufmanWinn13},
but without a clear resolution. 

Sparked by the new space-based era of high-precision time-series photometry,
a recent surge of results from the asteroseismology of red giants has
shown that detailed and highly precise information can be obtained about
these stars \citep[e.g.][]{Bedding11,Beck12,Mosser12c,Stello16a}.   
In particular, stellar mass can be measured 
\citep{Stello08,Kallinger10}, which makes asteroseismology an obvious
way to resolve the dispute about the retired A-star planet-host
masses. 

One retired A-star, the red-giant-branch star HD
185351, was observed by \kepler. 
The initial analysis by \citet{Johnson14} found different masses spanning
1.60-1.99\msol\ depending on the combination of seismic 
and interferometric input, as compared to 1.87\msol\ for the purely
spectroscopic-based mass.  The seismic inputs were \dnu, the frequency 
spacing between overtone modes, and \numax, the frequency of maximum power.  
Recent results including also the period spacing between dipole mixed modes 
(\dP) as an extra seismic quantity, and non-standard physics provided stronger 
constraints, indicating more definitively 
a lower mass of $1.58\pm 0.03\,$\msol\ for this star \citep{Hjoerringgaard17}.   

Even more recently, \citet{Campante17} used K2 observations of another
retired A-star, HD 212771, to measure its mass from
asteroseismology to be $1.45\pm 0.10\,$\msol.  Despite this star being in
a very similar evolutionary stage to HD 185351, the seismic mass of HD
212771 was larger than the original spectroscopy-based value of $1.15\pm
0.08\,$\msol\ \citep{Johnson10}.
However, subsequent spectroscopic investigations also estimated its mass to
be larger ($1.51\pm 0.08\,$\msol\ \citep{Mortier13}, $1.60\pm
0.13\,$\msol\ \citep{Jofre15}) than reported by \citet{Johnson10}.

With only two seismic targets showing inconclusive results, it is still not
clear if the stellar masses 
are {\it systematically} overestimated 
in the planet discovery papers, 
which would affect
conclusions about how planet occurrence rates depend on stellar mass,
and potentially lead to a different explanation for the high occurrence of
planets in the retired A-star
sample (currently attributed to the high stellar mass). 

In this paper, we investigate the retired A-star mass controversy by
observing eight planet-hosting red giants using the ground-based SONG
telescope to detect solar-like oscillations. This allows us to estimate the
stellar mass using asteroseismology, and hence to investigate
whether there is a general problem with the
previously adopted mass scale of the retired A-star planet hosts.

\section{Target selection and observations}\label{observations}
\begin{figure}
\includegraphics[width=\columnwidth]{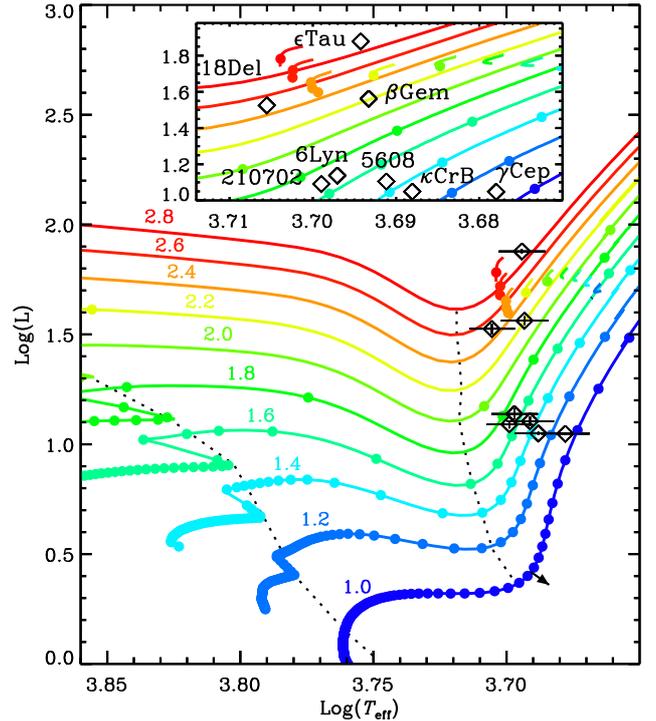}
\caption{HR-diagram showing MESA \citep{Paxton13} stellar
  evolution tracks of solar metallicity from \citet{Stello13}.  The
  likelihood of finding a star in a given state of evolution (for a given
  mass) is illustrated by the filled dots
  along each track, which are equally spaced by 50 million years in stellar age.
  Masses in solar units are shown.  The black arrow near the bottom of the
  1.0\msol\ red giant branch illustrates how much the tracks
  shift if [Fe/H] is increased by 0.2 dex.  Dotted fiducial lines are
  indicative of the transitions from the main sequence to subgiants, and
  from the rapidly cooling subgiants (at roughly constant radius) to the
  rapidly expanding red giants (at roughly constant \teff).  The
  planet-hosting targets are shown with diamonds and the adopted
  $1\sigma$-errorbars.  The helium-core burning 
  models are those within the range $1.6 \lesssim \log(L) \lesssim
  1.8$. The inset shows a close-up. 
\label{hrd}} 
\end{figure}

We note that the stars in question are often referred
to as ``subgiants'' following the historical spectroscopic classification that some of
them carry. 
However physically, they are either helium-core-burning (red-clump stars),
or burning hydrogen in a shell around an inert helium core with 
luminosities below the red clump but already in the red-giant-branch phase
(with radius, and hence luminosity increasing rapidly at roughly constant
\teff).  We therefore refer to them as red giants.  The mass controversy is
starkest for the red-giant-branch stars.  Their evolutionary speed is
highly mass-dependent, making it much more likely to find such stars with M  
$\lesssim 1.5\,$\msol\ than M $\gtrsim 1.5\,$\msol, as illustrated by the 
mass-dependent density of dots along the stellar evolution tracks in
Figure~\ref{hrd}. Genuine retired A-stars are expected to be rare. 

We selected our targets from the EOD\footnote{www.exoplanets.org}.  
The initial criteria were $3.75>\log($\teff/K$)>3.65$
($5623\,K >$ \teff\ $>4467\,K$) and $\log(L/$\lsol$)>0.75$ ($L>5.62\,$\lsol).
For this we used \teff\ from the EOD and derived
luminosity using the Hipparcos distance, $V$ magnitude, and a
metallicity-dependent bolometric correction \citep[ Eq. 18]{Alonso99},
ignoring extinction due the proximity of our targets. 
From this initial selection we chose the six brightest stars in the
northern sky with $\log g > 3$ and the two brightest stars with $\log g < 3$
(see Figure~\ref{hrd}). 

The time-resolved radial velocities were obtained with the
robotic 1-metre Hertzsprung SONG telescope on Tenerife \citep{Andersen14,Grundahl17}
during the period from August 2014 to December 2015 using its \'echelle
spectrograph.  Our strategy was to observe the stars long enough that
the frequency of maximum oscillation power, \numax, could be determined
from single-site observations to a precision of about 15\%.  This should
allow us to make conclusions about difference in mass between
seismology- and spectroscopy-based values in an ensemble sense even if
not on a single-star basis.  To determine the length of time-series
observations required, we used data of the red giant
$\xi\,$Hya obtained using the Coralie spectrograph on the 1.2-metre Euler
telescope at La Silla \citep{Frandsen02}, which has similar performance to
SONG.  These data comprised 30 full consecutive nights of observations
that clearly showed stellar oscillations with frequencies centred at
\numax\ $\sim 90$\muhz\ \citep{Stello04}.  By splitting the series into
multiple segments we found that the intrinsic \numax\ scatter across
segments reached about 15\% if the segments were 5--10 days long\footnote{The accuracy of \numax\ also reached about 15\%, 
measured as the average deviation of individual segment \numax\ values from the reference value based on the full
30-night dataset.}; this
guided the required minimum observation length per star.

The lengths of the observations varied between 5 and 13 nights, 
and the number of spectra for each target varied from night to night
due to constraints from weather, visibility, and the execution of other 
observing programs, which was not critical for our purpose of 
measuring \numax\ to within 15\%.  However, we do note that such short 
single-site data on red giants do not allow us to measure \dnu\ or 
individual mode frequencies.
The SONG \'echelle spectroscopy made use of an iodine cell
for high-precision wavelength calibration. Exposure times were tuned to
ensure sufficiently sampling of the oscillations while keeping the spectral
signal-to-noise ratio above $\sim$100 in the wavelength range with a large number
of iodine lines.   Table~\ref{tab1} lists the basic parameters for the 
data obtained for each target.

\begin{table}
{\footnotesize
\centering
\caption{Observing parameters for the targets.\label{tab1}}
\begin{tabular}{lccccccc}
\hline
 Star            & $V$  & T$_{\mathrm{exp}}$ &  N$_{\mathrm{exp}}$ & R     & N$_\mathrm{night}^{\mathrm{obs}}$ & N$_\mathrm{night}^{\mathrm{span}}$ &  $\sigma_\mathrm{RV}$ \\
                 &      & [s]             &                   &       &                                &                                &  [m/s]   \\
\hline                                                                                                                              
 $\epsilon\,$Tau*& 3.53 & 180             &   941             & 77k   & 8                              & 9                              &  2.76  \\  
 $\beta\,$Gem*   & 1.15 &  20             &  5410             & 77k   & 5                              & 6                              &  1.61  \\  
 18 Del*         & 5.51 & 600             &   358             & 77k   & 9                              & 11                             &  2.53  \\  
 $\gamma\,$Cep   & 3.21 & 120             &  1758             & 90k   & 13                             & 13                             &  2.00  \\
 HD 5608         & 6.00 & 600             &   171             & 77k   & 7                              & 9                              &  2.70  \\  
 $\kappa\,$CrB   & 4.79 & 300             &   495             & 90k   & 7                              & 13                             &  2.11  \\  
 6 Lyn           & 5.86 & 600             &   206             & 77k   & 5                              & 5                              &  3.12  \\  
 HD 210702       & 5.93 & 600             &   126             & 77k   & 5                              & 7                              &  2.52  \\  
\hline
\multicolumn{8}{l}{$V$: magnitude.}\\
\multicolumn{8}{l}{T$_{\mathrm{exp}}$: exposure time.}\\
\multicolumn{8}{l}{N$_{\mathrm{exp}}$: number of exposures.}\\
\multicolumn{8}{l}{R: spectrograph resolution.}\\
\multicolumn{8}{l}{N$_\mathrm{night}^{\mathrm{obs}}$: number of observing nights.}\\
\multicolumn{8}{l}{N$_\mathrm{night}^{\mathrm{span}}$: length of time series.}\\
\multicolumn{8}{l}{$\sigma_\mathrm{RV}$: mean radial-velocity precision.}\\
\multicolumn{8}{l}{* Most likely red clump stars (see Figure~\ref{hrd}); not in conflict with}\\
\multicolumn{8}{l}{the known planet orbits.}\\
\end{tabular}
}
\end{table}

\begin{figure*}
\includegraphics[width=17.6cm]{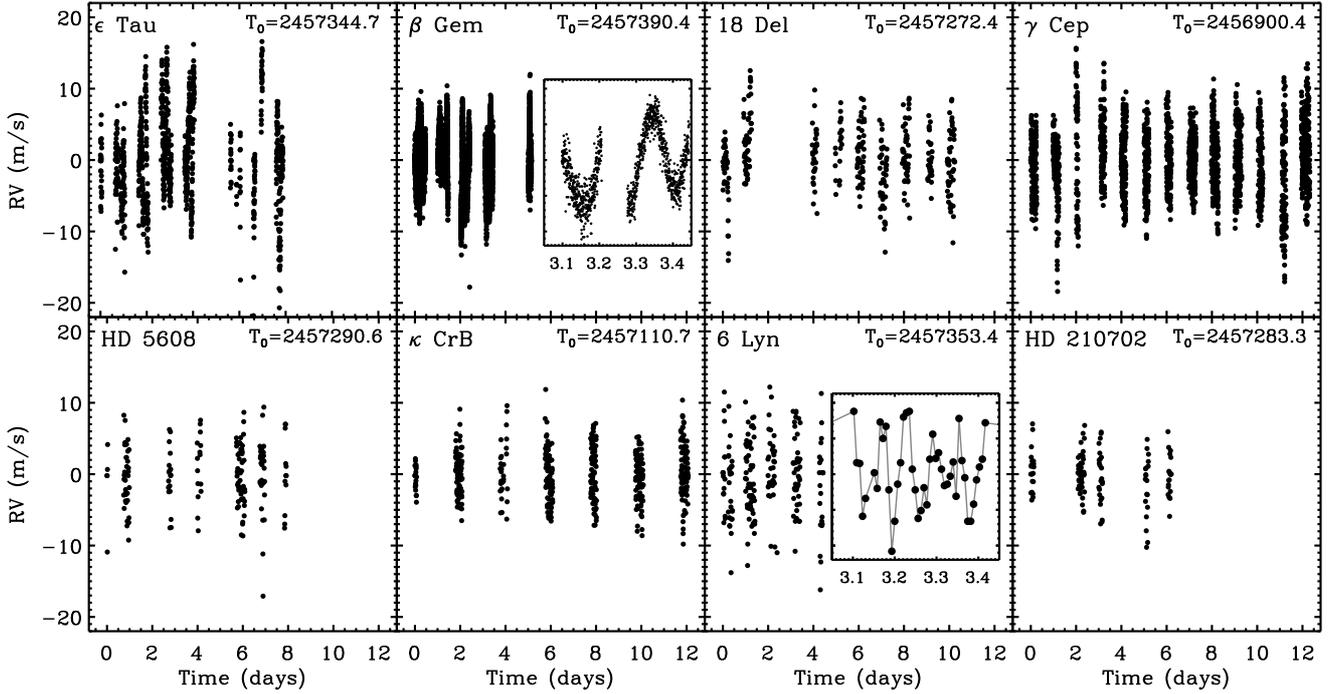}
\caption{Radial-velocity time series of the eight planet-hosting stars.  The
  time, T$_0$ (BJD) of the first data point is indicated. For
  the stars $\beta\,$Gem and 6 Lyn the inset shows a single night of observation.
\label{timeseries}} 
\end{figure*} 
The data reduction into 1-D spectra was performed using an extraction pipeline
based on the C++ implementation by \citet{Ritter14} of the IDL routines by
\citet{Piskunov02} for order-tracing and extraction of \'echelle spectra.  
The calculation of the radial-velocity time series was performed using the
iSONG software \citep{Antoci13,Grundahl17}, 
which follows the approach by
\citet{Butler96}.  We applied a post-processing high-pass filter with a
characteristic cut-off frequency of $\sim3$\muhz\ to remove any slow trends
in the data that could otherwise result in power leaking into the
frequency range of the stellar oscillations.  The final time series are
shown in Figure~\ref{timeseries}.  The radial-velocity scatter, which
typically ranges about $\pm 10\,$m/s, is dominated by the oscillations, as
illustrated by the insets for the stars $\beta\,$Gem and 6 Lyn.

\section{Stellar parameters}\label{parameters}
To make predictions of the expected seismic signal, \numax, we
used the empirical scaling relation by \citet{Brown91} and \citet{KjeldsenBedding95},
which relies on the assumption that \numax\ is proportional to the acoustic
cut-off frequency, hence:
\begin{eqnarray}
  \nu_\mathrm{max}/\nu_\mathrm{max,\odot} 
     &\simeq& \frac{M/\mathrm{M_\odot} }{ (R/\mathrm{R_\odot})^2 (T_\mathrm{eff}/T_\mathrm{eff,\odot})^{0.5} }\nonumber\\
     &=&      \frac{M/\mathrm{M_\odot} (T_\mathrm{eff}/T_\mathrm{eff,\odot})^{3.5}}{L/\mathrm{L_\odot}},
\label{scaling}
\end{eqnarray}
where $\nu_\mathrm{max,\odot}=3090\,$\muhz\ \citep{Huber09} and
$T_\mathrm{eff,\odot}=5777\,$K.  
This empirical relation has been verified
to be good to within at least 5\% for red giants of near solar metallicity
\citep[][ see also Section 4.1]{Huber12,Gaulme16}.
For each star, we therefore required estimates of mass, luminosity, and
effective temperature.  We adopted the spectroscopic-based masses from the
EOD 
(Table~\ref{tab2}, column 6), which are ultimately
those we want to compare with the seismology.  
We note that these masses have been updated compared to those disputed by
\citet{Lloyd11} (Table~\ref{tab2}, column 7), and we comment on those earlier 
mass results later.

To calculate luminosities, we used Hipparcos parallaxes (Table~\ref{tab2},
column 5), spectroscopic \teff\ and [Fe/H] from the 
EOD 
sourced from the same papers as the adopted mass to be self-consistent 
for the later comparison
(Table~\ref{tab2}, columns 3 and 4), and Tycho $V_{T}$ photometry
\citep{Hoeg00} as input to the direct method implemented in
\texttt{isoclassify} \citep{Huber17}\footnote{\url{https://github.com/danxhuber/isoclassify}}.   
In summary,
we sampled distances following a posterior calculated from the Hipparcos
parallax. 
The precise parallaxes make the posteriors
insensitive to the adopted prior.  For each distance sample, we calculated
the extinction, $A_V$, using the map by \citet{Green15}, as implemented in the
\texttt{mwdust} package by \citet{Bovy16}, and combined this with
independent random normal samples for the apparent magnitude and \teff\ to
calculate luminosities (Table~\ref{tab2}, column 8) and hence radii
(Table~\ref{tab2}, column 9).  Bolometric corrections were derived by
linearly interpolating \teff, $\log g$, [Fe/H], and $A_{V}$ in the MIST/C3K
grid (Conroy et al., in 
prep.\footnote{\url{http://waps.cfa.harvard.edu/MIST/model\_grids.html}}), 
but $\log g$ had little effect on the result (0.5 dex shifts changed the
correction by $0.006\,$mag).
The resulting distributions were used to calculate the mode and 1-$\sigma$
confidence interval for luminosities.  We also used the grid-modelling
method in \texttt{isoclassify} as an alternative approach, which allowed us to
fit for reddening.  This yielded consistent results to the direct method.
Finally, we derived the predicted \numax\ values listed in Table~\ref{tab2}
(column 10).
\begin{table*}
{\footnotesize
\centering
\caption{Observed parameters of the planet-hosting targets.\label{tab2}}
\begin{tabular}{l|cccccc|ccc|ccc}
\hline\hline
                & \multicolumn{6}{c}{Literature}                                                  & \multicolumn{3}{c}{Derived}                     & \multicolumn{3}{c}{Asteroseismology}  \\
 Star           & $\log g$  & \teff      & [Fe/H]      & $\pi$      & $M$        & $M_\mathrm{old}$ & $L$          & $R$       & $\nu_\mathrm{max,pre}$ & $\nu_\mathrm{max,obs}$ & $\log g$ & $M$      \\
                & [dex]$^1$ & [K]$^1$    & [dex]$^1$   & [mas]$^2$  & [\msol]$^1$ & [\msol]$^3$     & [\lsol]      & [\rsol]   & [\muhz]               & [\muhz]$^{8}$           & [dex]    & [\msol]  \\
          (1)   & (2)       & (3)        & (4)         & (5)        & (6)         & (7)            & (8)          & (9)       & (10)                  & (11)                  & (12)     & (13)     \\
\hline                                                                                                                                                                                      
$\epsilon\,$Tau & 2.62(15)  & 4746(70)   & 0.17(6)     & 22.24(25)  & 2.73(10)    & 2.70$^4$        &  75.54(1.80) & 11.8(5)  & 64.8(5.4)             & 56.9(8.5)             & 2.67(8)  & 2.40(36) \\ 
$\beta\,$Gem    & 2.91(13)  & 4935(49)   & 0.09(4)     & 96.54(27)  & 2.08(9)     & 1.86$^4$        &  36.50(1.69) & 8.21(37) & 101(10)               & 84.5(12.7)            & 2.84(8)  & 1.73(27) \\ 
18 Del          & 3.08(10)  & 5076(38)   & 0.0(?)      & 13.28(31)  & 2.33(5)     & 2.30$^5$        &  33.52(1.77) & 7.51(34) & 137(12)               & 112(17)               & 2.97(9)  & 1.92(30) \\
$\gamma\,$Cep   & 3.10(27)  & 4764(122)  & 0.13(6)     & 70.91(40)  & 1.26(14)    & 1.59$^4$        &  11.17(16)   & 4.88(22) & 177(24)               & 185(28)               & 3.17(8)  & 1.32(20) \\ 
HD 5608         & 3.25(16)  & 4911(51)   & 0.12(3)     & 17.74(40)  & 1.66(8)     & 1.55$^6$        &  12.74(62)   & 4.89(23) & 228(23)               & 181(27)               & 3.17(8)  & 1.32(21) \\ 
$\kappa\,$CrB   & 3.15(14)  & 4876(46)   & 0.13(3)     & 32.79(21)  & 1.58(8)     & 1.80$^7$        &  11.20(17)   & 4.70(20) & 241(21)               & 213(32)               & 3.24(8)  & 1.40(21) \\ 
6 Lyn           & 3.16(5)   & 4978(18)   &-0.13(2)     & 17.92(47)  & 1.82(13)    & 1.82$^{5}$      &  13.74(73)   & 5.01(25) & 243(28)               & 183(27)               & 3.18(9)  & 1.37(22) \\ 
HD 210702       & 3.36(8)   & 5000(44)   & 0.04(3)     & 18.20(39)  & 1.71(6)     & 1.85$^{4}$      &  12.33(52)   & 4.68(22) & 258(23)               & 223(33)               & 3.26(9)  & 1.47(23) \\ 
\hline
 \multicolumn{13}{l}{$^1$ Source: EOD (exoplanets.org), which refers to \citet{Mortier13} except for 6 Lyn for which it is \citet{Sato08} ($\log g$, \teff, and}\\
   \multicolumn{13}{l}{[Fe/H]) and \citet{Bowler10} ($M$).  Although the quoted uncertainties are typically below 50K (\teff) and 0.04dex ([Fe/H]), we assume}\\
   \multicolumn{13}{l}{$\sigma_{T_\mathrm{eff}}=100\,$K and $\sigma_\mathrm{[Fe/H]}=0.1\,$dex to derive columns 8--10 and 12--13.}\\  
 \multicolumn{13}{l}{$^2$ Source: Hipparcos \citep{Leeuwen07}. We note that HD 5608
 also has a TGAS parallax of 17.13(33)mas}\\  
   \multicolumn{13}{l}{\citep{Prusti16}, which would push its inferred (spectroscopic and seismic) masses up by about 0.1\msol.}\\
 \multicolumn{13}{l}{$^3$ Masses in the EOD before the \citet{Mortier13} updates, and originally disputed by \citet{Lloyd11}.}\\
 \multicolumn{13}{l}{$^4$ \citet{Johnson07a}; HD 5608 and $\gamma\,$Cep were not part of the original disputed set.}\\
 \multicolumn{13}{l}{$^5$ \citet{Bowler10}}\\
 \multicolumn{13}{l}{$^6$ \citet{Sato12}}\\
 \multicolumn{13}{l}{$^7$ \citet{Johnson08}}\\
 \multicolumn{13}{l}{$^8$ We adopt an uncertainty of 15\% in $\nu_\mathrm{max,obs}$ (Sect.~\ref{analysis}).}\\
 \multicolumn{13}{l}{Note: Uncertainties are shown in compact bracket form:
   e.g. $2.35(5)=2.35\pm 0.05$, $2.35(15)=2.35\pm 0.15$, $15.6(1.3)=15.6\pm 1.3$.}\\
\end{tabular}
}
\end{table*}

\section{Seismic data analysis}\label{analysis}
\begin{figure*}
\includegraphics[width=17.6cm]{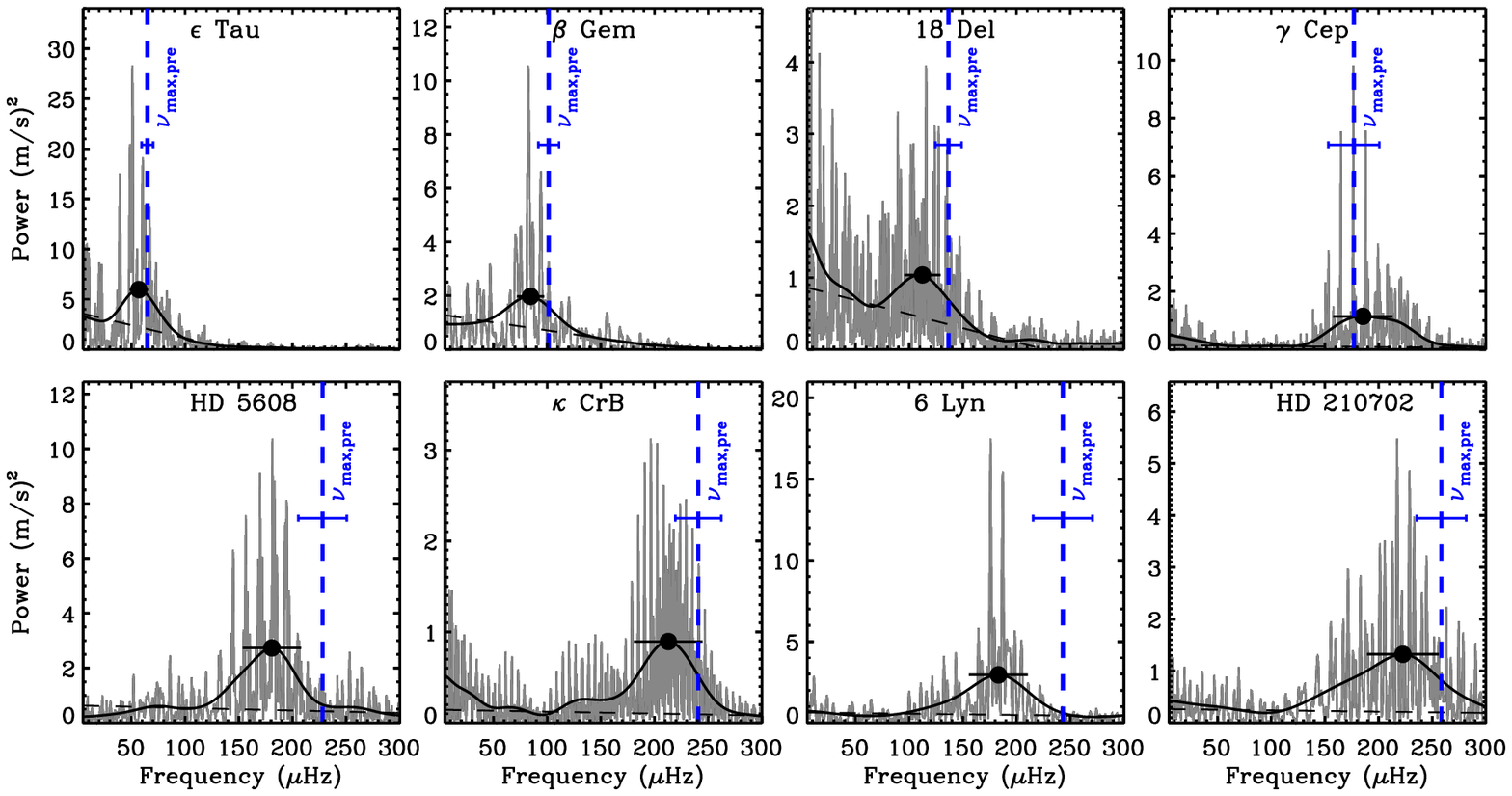}
\caption{Power spectra of the eight planet-hosting stars (ordered by
  \numax).  The smoothed
  spectra (black curves) and the location of 
  the observed \numax\ are shown (large dot), including $1\sigma$
  errorbars.  The vertical blue dashed lines indicate the predicted
  \numax\ assuming the stellar mass in the EOD. 
  The uncertainty on its location is shown by the blue errorbar.
  The dashed black line shows the noise 
  level in the region around the oscillations.  
\label{powerspectra}} 
\end{figure*}

The SONG power spectra of our stars are shown in
Figure~\ref{powerspectra} (grey curves); all showing clear excess power from
oscillations.  As expected the granulation background at low frequencies from
velocity measurements is much lower than is typically seen from photometry
\citep[][ their Fig.2]{Stello15}. 
Also, because these are short single-site radial-velocity observations our 
\kepler\ pipeline \citep{Huber09} for measuring \numax\ is not suited for these data.  
Nevertheless, we use the same basic approach as described in \citet{Huber09} 
for locating \numax, and an approach similar to that by \citet{MosserAppourchaux09} for 
measuring and subtracting the noise.  The location of \numax\ (large 
black dot), is found as the highest point of the heavily smoothed spectra 
(black curves) 
after subtracting the background noise estimated by a linear fit to the noise 
on either side of the oscillation power. 
However, we stress that ignoring the slope in noise did not change the 
\numax\ estimate significantly (below 1\%), suggesting that our adopted method 
for measuring the noise does not affect our conclusions.
The observed \numax\ values are listed in Table~\ref{tab2} (column 11).  We
adopted an uncertainty of 15\% on $\nu_\mathrm{max,obs}$ according to our
derivation from the $\xi\,$Hya observations that was based on the 
scatter between independent short time series (Sect.\ref{observations}).  
We regard this as a conservative estimate because it accounts for the
systematic uncertainty arising from the stochastic nature of the oscillations, 
which can cause the power excess on a single-star basis to be skewed 
differently from epoch to epoch.  This systematic uncertainty is much larger 
than the statistical uncertainty in measuring \numax\ on a single data set 
when the length of the time series is comparable to, or shorter than, the mode 
lifetime (as in our case) \citep{Dupret09,Corsaro15}.  In comparison to our 
adopted conservative uncertainty of 15\%, we note that $\gamma\,$Cep has roughly 
60 consecutive nights of observation from another SONG program (Palle et al. 
in prep.), which shows only 5\% scatter in \numax\ across 5--10-day segments. 
From $\nu_\mathrm{max,obs}$ and Eq.~\ref{scaling}, we derived the  
asteroseismic $\log g$ (column 12), using \teff\ (column 3) and derived the 
seismic mass (column 13) using also $L/$\lsol\ (column 8).
We remind the reader that becase our time series are short and single-site observations, 
we cannot determine individual frequencies or \dnu, in order to get additional 
seismic mass diagnostics.

It is evident from Figure~\ref{powerspectra} that all but one star show the
oscillation power centred below the predicted 
\numax\ shown by the blue vertical dashed lines.  
Individually, they would all be regarded `in agreement' with the
observations at the $2\sigma$-level, but as an ensemble maybe not so.
Under the assumption that the predicted \numax\ values are equal to the true
values, the chance of observing a lower \numax\ in seven out of eight stars is
only $\sim 3\,$\%, 
without taking into account the magnitude of the difference between predicted
and observed values.  However, 
a matched $t$-test across the ensemble shows the absolute
differences to be highly significant; we can reject the H0 hypothesis 
(that the predicted and observed values have a common average)
at the 0.5\% level (1.1\% if we had ignored
the likely clump stars, Table~\ref{tab1}). 
Hence, this shows a systematic overestimation of the predicted \numax. 
The ratio between the predicted and observed \numax\ ranges from 0.95 to 1.33
across the sample, with an average of $1.17\pm 0.04$
(the same if we had ignored clump stars).  If this difference is entirely
due to the adopted stellar mass for predicting \numax, it suggests that the
previously published spectroscopic-based masses that we adopted for these 
stars were generally overestimated by that factor (compare 
Table~\ref{tab2} column 6 and 13).  Interestingly, the star that agrees 
best with the seismology ($\gamma\,$Cep), has a dynamic mass of 
$1.40\pm 0.12\,$\msol\ \citep{Neuhauser07},
and was not included in the retired A-star sample by \citet{Johnson07a}. 
We note that, on average, the spectroscopic-to-seismic mass ratio would have 
been 1.24, if we had adopted the masses (Table~\ref{tab2}, column 7)  
and associated spectroscopic \teff\ and [Fe/H] that were in the EOD at the 
time they were disputed by \cite{Lloyd11}.
It is also noticeable that the scatter between the two sets of
spectroscopic masses (Table~\ref{tab2}, columns 6 and 7) is larger than
suggested by their formal uncertainties.
In the following, we look into which factors other than adopted mass could
make our predicted \numax\ consistently too large.  

\subsection{Potential systematics}
If our adopted temperature scale is off, it would affect all parameters
that go into predicting \numax\ (Eq.~\ref{scaling}).  Re-running
\texttt{isoclassify} with $100\,$K cooler \teff\ input results in an
estimated luminosity increase of 4\%, a mass decrease of 2--3\%, in addition
to the 7\% decrease in \teff$^{3.5}$, which all combined decreases the
estimated \numax\ by 14\%\footnote{Here, we assumed we could approximate 
the effect from such a \teff\ shift on the adopted spectroscopic-based mass 
by the mass change seen when running \texttt{isoclassify} in the grid-based 
mode with two \teff\ scales essentially replicating the typical 
spectroscopic-based approach for estimating stellar mass.}. 
Hence, the predicted \numax\ would agree with the observations if 
our adopted \teff\ scale was too hot by 100--150$\,$K.  There is indeed 
large scatter in \teff\ for these stars in the literature (see Simbad), 
but little empirical evidence to which \teff\ scale is the most correct
one for these stars.  The most fundamental test of \teff\ comes from 
interferometry.  For the three stars in common with our sample, we 
compared our adopted spectroscopic \teff\ with the scale found
from interferometric angular diameters and bolometric fluxes by White et al. 
(in prep.).  It shows that our adopted \teff\
is the same for 6 Lyn, 9K hotter for $\kappa\,$CrB, and 50K hotter for HD
210702, suggesting our \teff\ scale is not too hot at the 100--150$\,$K level, 
and hence does not support the notion that the discrepancy in 
Figure~\ref{powerspectra} is caused by our \teff\ scale being too hot.  
We note that there is some tension between interferometric angular
diameters measured from different instruments and thus the adopted \teff\ 
scale \citep[e.g.][ and references therein]{Casagrande14, Huber17}, but
the higher spatial resolution optical interferometry by White et 
al. should be less affected by systematic errors than previously
published values. 

Turning our attention to the adopted metallicities, we see significant
scatter across literature values.  As noted earlier, we tried to compensate
by adopting a larger metallicity uncertainty than quoted in
Table~\ref{tab2}. 
However, a systematic shift in metallicity of
+0.1 dex would change our predicted \numax\ by about 1\% for clump stars
and 4\% for red-giant-branch stars; a change totally dominated by the
change in the adopted spectroscopic-based mass.  Again, this is assessed
using \texttt{isoclassify} in its grid-based mode.  This mass-metallicity
dependence is illustrated by the small black arrow in Figure~\ref{hrd}
(lower-right). 
Although there are no indications of the adopted [Fe/H] being
systematically off, such metallicity systematics could only play a minor
role in loosening the tension between the predicted and observed \numax.

Finally, could the \numax\ scaling relation be systematically off by
15--20\% for red giants, resulting in over-prediction of \numax?
The most direct test of this relation for red giants was carried out by
\citet{Gaulme16}. They used oscillating giants in eclipsing binaries, to
measure a dynamic $\log g$, and hence \numax\ given \teff, totally independent of
seismology.  Comparing that with the seismically measured \numax\ showed
agreement within 3-4\% (their Fig.7). We note that their sample
comprised generally more evolved stars (lower $\log g$) than ours and the 
larger $\log g$ stars showed the lowest discrepancies.
\citet{Huber12} used interferometry to obtain independent
measurements of stellar radius, which combined with the relation
\dnu\ $\simeq M^{0.5}R^{-1.5}$ (all in solar units) provides mass
and hence an expected \numax, given \teff.  
Although the uncertainties on their scaled \numax\ values were relatively
large, their result (their Fig.8) rules out a systematic error at
the 15--20\% level required to explain the \numax\ difference seen in
Figure~\ref{powerspectra}.  
Most other studies attempting to verify the seismic-inferred mass from
scaling, use the relation $M \simeq
\,$\numax$^3$\dnu$^{-4}$\teff$^{1.5}$ (all in solar units), which has a much
stronger dependence on \numax\ than Eq.~\ref{scaling}, and a high
dependence on \dnu, and their results are hence not directly applicable to
our case.  
In summary, it seems unlikely that Eq.~\ref{scaling} is off by
15--20\%.  
However, we note that additional confirmation of the \numax\ 
relation will have to wait till the Gaia DR2 results are published, in 
addition to more interferometric measurements of red giants.

\section{Conclusion}
We used radial-velocity time series from the ground-based SONG
telescope to determine the asteroseismic masses of eight planet-hosting
red giants (`retired A-stars'). Three are possibly helium core-burning
clump stars, while the rest are unambiguously in the ascending
red-giant-branch phase.  While our observations are too short to firmly 
establish the mass with high precision for individual stars, our sample 
is large enough to make conclusions on the ensemble.  Based on our reported 
systematic offset between predicted and observed \numax\ of 15--20\%,
the results indicate that the previous mass determinations adopted here, 
which were based purely on spectroscopic constraints, are on average 
overestimated by about 15--20\% for these evolved stars, at least those 
with $M_\mathrm{spec}\gtrsim 1.6$\msol.  This conclusion assumes that 
potential systematics from the adopted \teff\ scale and the \numax\ scaling 
relation are negligible.  Based on our findings, these potential 
systematics could conspire and add up to a 4--5\% contribution of the 
observed offset. 
Our result seems consistent with the offset found by \citet{Hjoerringgaard17} 
for the $M_\mathrm{spec}\sim 1.9$\msol\ giant HD 185351, and the lower
offset (though within $1\sigma$) found by \citet{Campante17} for the
$M_\mathrm{spec}\sim 1.5$\msol\ giant HD 212771 compared to
\citet{Mortier13} (our main source of spectroscopic-based results).  
Our results also seem compatible with \citet{North17} who find no
mass offset on average for their generally lower-mass sample of red giants.

From 2018, many of the evolved planet hosts will be observed by TESS for at
least one month \citep{Ricker14}. From these data, we can expect to 
measure \numax, and probably \dnu, and for those in the TESS continuous
viewing zones, we should be able to also measure \dP\ of the g-modes, which
provides additional constraints to the modelling
\citep{Hjoerringgaard17}. These investigations will be further enhanced by
including Gaia DR2 parallax measurements into the mass estimates, reducing
observational uncertainties that will enable precise mass estimates on each
single star, and not just the ensemble.  Gaia DR2 will also provide
confirmation of the \numax\ scaling relation independent of those 
already at hand from eclipsing binaries and interferometry.



\section*{Acknowledgments}
This research was based on observations made with the Hertzsprung SONG telescope operated
on the Spanish Observatorio del Teide on the island of Tenerife by the Aarhus
and Copenhagen Universities and by the Instituto de Astrofísica de Canarias.
We would like to acknowledge the Villum Foundation, The Danish Council for Independent
Research | Natural Science, and the Carlsberg Foundation for their support in building the
SONG prototype on Tenerife. Funding for the Stellar Astrophysics Centre is
provided by The Danish National Research Foundation (Grant DNRF106). The
research was supported by the ASTERISK project (ASTERoseismic
Investigations with SONG and Kepler) funded by the European Research
Council (Grant agreement no.: 267864). 
D.S. acknowledges support from the Australian Research Council.
PLP acknowledges the support of the Spanish Ministry of Economy and
Competitiveness (MINECO) under grant AYA2016-76378-P.
L.C. gratefully acknowledges support from the Australian Research Council (grants DP150100250, FT160100402).
This research has made use of the Exoplanet Orbit Database
and the Exoplanet Data Explorer at exoplanets.org.
We thank Joel Zinn for helpful discussion.




\bibliographystyle{mnras}
\bibliography{bib_complete} 








\bsp	
\label{lastpage}
\end{document}